# Plasmon-Molecule Remote Coupling via Column-Structured Silica Layer for Enhancing Biophotonic Analysis

Takeo Minamikawa,[1,2,3,*,§] Reiko Sakaguchi,[4] Yoshinori Harada,[2] Hideharu Hase,[5] Yasuo Mori,[4,5] Tetsuro Takamatsu,[2,6] Yu Yamasaki,[7] Yukihiro Morimoto,[7,8] Masahiro Kawasaki,[9] and Mitsuo Kawasaki[9, §]

[1]Division of Interdisciplinary Researches for Medicine and Photonics, Institute of Post-LED Photonics, Tokushima University, Tokushima 770-8506, Japan.

[2]Department of Pathology and Cell Regulation, Graduate School of Medical Science, Kyoto Prefectural University of Medicine, Kyoto 602-8566, Japan.

[3]PRESTO, Japan Science and Technology Agency (JST), Tokushima 770-8506, Japan.

[4]Institute for Integrated Cell-Material Sciences (iCeMS), Kyoto University, Kyoto 606-8501, Japan.

[5]Department of Synthetic Chemistry and Biological Chemistry, Graduate School of Engineering, Kyoto University, Kyoto 615-8510, Japan.

[6]Department of Medical Photonics, Graduate School of Medical Science, Kyoto Prefectural University of Medicine, Kyoto 602-8566, Japan.

[7]Technology & Engineering Division, Ushio Inc., Hyogo 671-0224, Japan.

[8]The Institute of Science and Industrial Research, Osaka University, Osaka 567-0047, Japan.

[9]Department of Molecular Engineering, Graduate School of Engineering, Kyoto University, Kyoto 615-8510, Japan.

**ABSTRACT**

We demonstrated remote plasmonic enhancement (RPE) by a dense random array of Ag nanoislands (AgNIs) that were partially gold-alloyed and attached with column-structured silica (CSS) overlayer of more than 100 nm in thickness. The physical and chemical protection of the CSS layer could lead to reducing the mutual impact between analyte molecules and metal nanostructures. RPE plate was realized just by sputtering and chemical immersion processes, resulting in high productivity. We found a significant enhancement on the order of $10^7$-fold for Raman scattering and $10^2$-fold for fluorescence by RPE even without the proximity of metal nanostructures and analyte molecules. We confirmed the feasibility of RPE for biophotonic analysis. RPE worked for dye molecules in cells cultured on the CSS layer, enabling the enhanced fluorescence biosensing of intracellular signaling dynamics in HeLa cells. RPE also worked for biological tissues, enhancing Raman histological imaging of esophagus tissues with esophageal adventitia of a Wistar rat attached atop the CSS layer. We also investigated the wavelength dependency of RPE on the on- or off-resonant with the dye molecular transition dipoles with various molecular concentrations. The results suggested that the RPE occurred by remote resonant coupling between the localized surface plasmon of AgNIs and the molecular transition dipole of the analyte via the CSS structure. The RPE plate affords practical advantages for potential biophotonic analyses such as high productivity and biocompatibility. We thus anticipate that RPE will advance to versatile analytical tools in chemistry, biology, and medicine.

## 1. INTRODUCTION

Fluorescence and Raman spectroscopy are essential analytical tools in biological and biochemical studies.[1-8] In fluorescence biosensing of live cells, various fluorescence probes offer information about intracellular metabolism and signaling mechanisms.[9-12] As for Raman spectroscopic biosensing of biological tissue, there has been considerable interest in the potential clinical application.[13-20] Currently available biophotonic methods are subject to some technical limitations. For fluorescence biosensing, since the sensor molecule is an additive that may disturb the natural metabolism in live cells, its dosage must be kept low, which causes a tradeoff between external perturbation and analytical sensitivity. The sensitivity problem could be serious in the case of tissue Raman imaging because intense and prolonged laser irradiation necessary for signal acquisition might induce photodamaging to specimens.

Plasmon-enhanced fluorescence and Raman spectroscopy offer a promising solution for analytes with weak optical signals. Surface-enhanced Raman scattering (SERS) and surface-enhanced fluorescence (SEFL) have many important applications in biophotonic analyses.[21-30] Nevertheless, the application of localized surface plasmon resonance (LSPR) phenomena to bioanalysis has been limited because the nanostructured metal surfaces need to be in proximity to the analyte molecules, possibly leading to the mutual degradation between analyte molecules and metal nanostructures.

In this paper, we report plasmon-mediated long-range enhancement of fluorescence and Raman scattering via dielectric nanostructures, namely, remote plasmonic enhancement (RPE). The key element for the present RPE function is the column-structured silica (CSS) layer that protects the mutual adverse impact between a dense random array of Ag nanoislands (AgNIs) and analyte molecules. The experimental evidence presented in this paper establishes the proposed concept of RPE. We have also demonstrated RPE-enhanced fluorescence biosensing for live cells and RPE-enhanced Raman imaging of biological tissues as practical validation of RPE.

## 2. EXPERIMENTAL SECTION

### 2.1 Materials

Most of the chemicals of special reagent grade, including basic fuchsine (FUC) dye, were used as received from Wako Pure Chemical Industries. Polyvinyl alcohol (PVA) was of polymerization degree of 500 or less and a saponification degree of 86–90%. Rhodamine 6G dye (R6G) was obtained from Exciton as a chloride salt. Intracellular $Ca^{2+}$ indicator fluo3-AM was from Dojindo Laboratories. An extracellular matrix (Matrigel) was from Corning. The frozen section compound (FSC 22) was from Leica Biosystems.

### 2.2 Preparation of RPE Plates

A dense array of AgNIs was grown by direct-current $Ar^+$ ion sputtering onto the smoother side of a float slide glass plate (Type S7213; Matsunami) in an apparatus similar to that used elsewhere.[31] A glow discharge at a negative voltage of 1.4 kV or less supplied to the cathode as an Ag target and a discharge current of 15 mA produced a dense random array of AgNIs in 5 min. The deposition of a CSS layer on the AgNIs layer was conducted using radio-frequency sputtering (model RFS-200; Ulvac) at an Ar pressure of 1.0 Pa. The radio-frequency power was adjusted to 100 W, at which we gained a $SiO_2$ deposition rate of 10 nm/min. The high-energy discharge plasma caused the substrate temperature up to 160 °C.

### 2.3 Structural and optical characterizations of RPE plates

Structural and optical characterizations of the RPE plates were made by field-emission scanning electron microscopy (FE-SEM; SU8000 and SU9000; Hitachi), atomic-force microscopy (AFM; VN-8000; Keyence), spectroscopic ellipsometry (FE-5000; Otsuka Electronics), and UV-VIS spectroscopy (UV-3600; Shimadzu).

**2.4 Gold(I)/Halide Bath Treatment on AgNIs**

The gold latensification has long been used in the silver halide photographic technology, which renders the Ag latent image a higher catalytic ability in photographic development.[32-33] This function stems from a nanoalloying of an $(Ag)_n$ latent image center into $(Ag)_{n-x}(Au)_x$ by replacement or into $(Ag)_n(Au)_m$ by plating. The standard protocol for the preparation of the gold(I)thiocyanate latensification solution is as follows; 0.5 g of KSCN is added to 40 mL of 0.1 wt% $NaAuCl_4 \cdot 2H_2O$ solution, is heated to boiling for a few minutes, 0.6 g of NaBr is added, and dilutes with water to 1 L. In the present study, the optimum solution was prepared by diluting this standard solution with 0.2 M NaBr by 30 times. We thus obtained a gold latensification bath containing $Au(SCN)_2^-$ at a concentration of 3.3 μM. The CSS-protected AgNIs were processed for 1–2 min at 65 °C, quickly rinsed with water, immersed in 0.2 M NaCl solution for 5 min at 75 °C, rinsed again with water, and dried under air conditions. The follow-up processing in 0.2 M NaCl solution at 75 °C led to substantial uptake of NaCl along the intercolumn boundaries.

**2.5 Spectroscopic Measurements**

A large body of the fluorescence and Raman spectra were measured using a simple laboratory-made spectrometer. It comprised a low-power He-Ne laser (633 nm, 0.7 mW) or a green diode laser (532 nm, 3 mW), two lenses in series to collect the emission from the sample on the RPE plate onto an aperture of a light receiver head, connected via fiber optics to the cooled multichannel analyzer (PMA-11 and C5966-31; Hamamatsu Photonics). The other one, a commercial portable spectrometer, was a coaxial probe Raman spectrometer (Raman-Probe-785; StellarNet) operated at the excitation wavelength of 785 nm at a high laser power of 600 mW.

High-resolution Raman spectra for the excitation laser wavelength of 633 nm were acquired with a Raman spectrometer (DXR3 SmartRaman Spectrometer; ThermoFisher Scientific). High-resolution Raman spectroscopy and imaging were conducted using a line illumination confocal Raman microscope (Raman-11; Nanophoton) with the excitation laser operating at 532 nm.

Fluorescence images of the cells were captured by an electron-multiplying charge-coupled device camera (ImagEM; Hamamatsu Photonics) operated with MetaFluor software (Molecular Devices). The excitation wavelength was 480 nm.

**2.6 Cultivation of Cells on RPE Plates and Incorporation of Fluorescent Biosensors**

HeLa cells were obtained from ATCC and were cultured routinely with Dulbecco's modified Eagle medium containing 10% fetal bovine serum, 30 units/mL penicillin, and 30 μg/mL streptomycin under 95% air, 5% $CO_2$ atmosphere at 37 °C. For $Ca^{2+}$ measurement in HeLa cells, they were trypsinized and $5\times10^4$ cells were plated onto a Matrigel-coated RPE plate or glass. After 2–6 h, cells were loaded with various concentrations of fluo3-AM at 37 °C for 30 min and washed with Tyrode's solution containing the following: 140 mM NaCl, 5 mM KCl, 1 mM $MgCl_2$, 2 mM $CaCl_2$, 10 mM glucose, and 10 mM HEPES (pH adjusted to 7.4 with NaOH). Histamine was applied to evoke intracellular calcium oscillation. The fluo3-AM fluorescence was measured in Tyrode's solution at ambient temperature.

## 2.7 Preparation of Biological Tissues

All animal experiments were conducted with the approval of and in accordance with guidelines from the Committee for Animal Research, Kyoto Prefectural University of Medicine (Permission No. M25-109). The esophagus with esophageal adventitia was excised from a Wistar rat after euthanasia. The esophagus with esophageal adventitia was immediately embedded in the frozen section compound, snapfrozen in dry ice-acetone, and stored at −80°C until cryostat sectioning. The frozen samples were sliced into 5 μm in thickness using a cryostat microtome (CM1900; Leica) and mounted without any fixation on an RPE plate or a 0.17-mm thickness cover glass (No.1; Matsunami).

## 3. RESULTS AND DICSUSSION

### 3.1 Key Structural Elements of RPE Plate

Figure 1a shows a two-dimensional dense random array of AgNIs, which we used as the basal layer constituting an RPE plate. Each AgNI measured 50–150 nm in lateral dimension and less than 20 nm in height. Unless otherwise protected, such AgNIs are highly vulnerable that they do not even withstand aerobic corrosion and undergo immediate corrosive dissolution in saline, i.e., under a common aqueous environment for biological experiments. The CSS layer, ~100 nm or more in thickness, worked as a robust protection layer for the AgNIs. Figure 1b shows a cross-sectional FE-SEM image of a 120 nm-thick CSS layer, in which a column structure with directional intercolumn boundaries is resolved. The intercolumn boundaries originate predominantly from the interstitial positions of the respective AgNIs. The spectroscopic ellipsometry measurement gave silica layer thicknesses that agreed with the FE-SEM images. The corresponding refractive index of 1.42 at the wavelength of 600 nm suggested a dense silica network comparable to that of fused silica with the refractive index of 1.46.

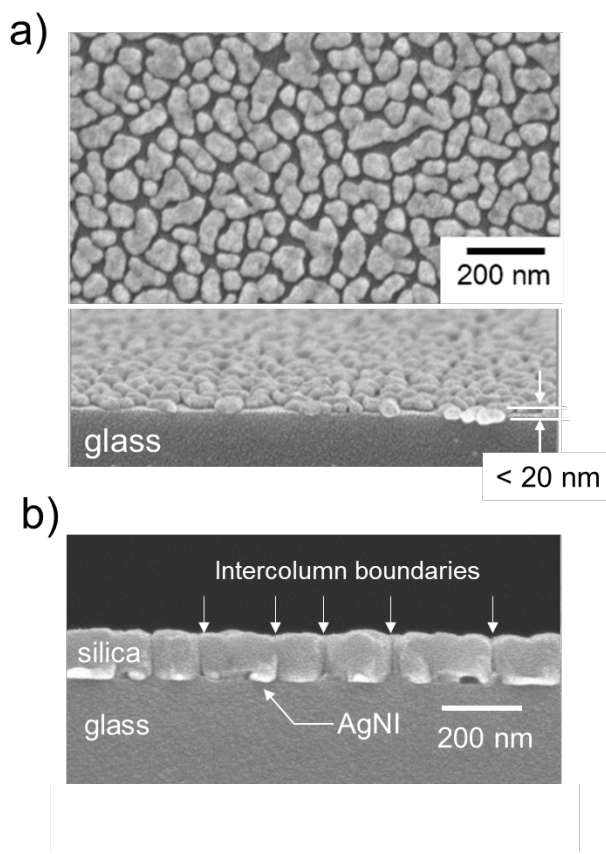

**Figure 1.** FE-SEM image of an RPE plate. (a) Top-view and an oblique-angle cross-sectional FE-SEM image showing a dense random array of AgNIs, 50–150 nm across and less than 20 nm in height. (b) A cross-sectional FE-SEM image of 120-nm thick CSS layer protected AgNIs, resolving the column structure characteristic of the sputter-grown silica layer over the AgNIs.

**3.2 Fundamental Confirmation of RPE**

AgNIs protected with the as-grown CSS layer already acquired the capability to bring significant Raman enhancement for various molecules, such as R6G, 2-naphthalene thiol, 1-hexadecane thiol, and cysteamine, adhered on top of the CSS layer, as shown in Figure 2. This pristine plate was not treated in the gold(I)/halide bath. The RPE-enhanced Raman spectrum of a spin-coated R6G in Figure 2a, compared to a spectrum of a 1.25 mM ethanolic solution of R6G on a slide glass ($8\times10^{16}$ molecules/cm² in scattering volume, see Figure S1), suggested that the corresponding Raman enhancement factor (EF) typically reached the order of $10^6$, which was comparable to that achieved in the conventional SERS. In practice, the RPE functioned not just for molecules adhered to the CSS layer, as shown in Figure 2, but also for various analyte species embedded or dispersed in certain polymers (typically PVA) laid on the CSS protection layer.

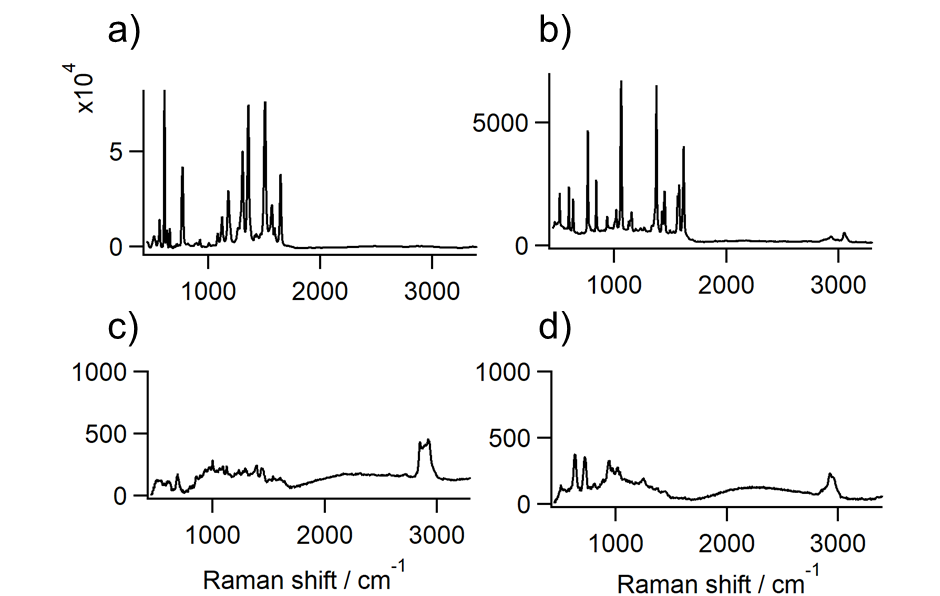

**Figure 2.** Enhanced Raman spectra of various molecules onto the pristine RPE plate without the gold(I)/halide bath treatment at 633-nm excitation. These molecules were spin-coated on the pristine RPE plate with the molecular concentration of 1–2 mM in ethanolic solutions, which corresponds to the molecular coverage on the order of $10^{14}$ molecules/cm². Enhanced Raman spectra of (a) R6G, (b) 2-naphthalene thiol, (c) 1-hexadecane thiol, and (d) cysteamine by the RPE plate. The reference spectra taken for the same series of spin-coated molecules on a slide glass gave no measurable Raman signals.

As for molecular species that were weakly bound to the CSS surface, they were readily washed out from the surface, and thereby the enhanced Raman signals were extinguished concomitantly. This proved that the observed enhancement occurred on the Raman-active species separated from the AgNIs by a more than 100 nm-thick CSS layer. For further proof of this long-range Raman enhancement, we also carried out an adhesive tape test (Figure S2), where we lightly dispersed 2-naphthalene thiol fine powders onto the adhesive side of a tape. We obtained enhanced Raman signals of 2-naphthalene thiol only when the tape was attached to the CSS surface of the RPE plate, and besides, the signals reversibly appeared and disappeared with the tap on and off the surface.

Furthermore, we found that the RPE function was upgraded by the gold(I)/halide bath treatment of the RPE plate. Figure 3 shows the gold(I)/halide bath treatment and the resultant extra broadening of the plasmon band in the near-infrared region. The effect of the gold(I)/halide bath treatment on enhanced Raman and fluorescence spectroscopies was observed by using FUC molecules embedded in a PVA layer attached to the CSS layer of an RPE plate, as shown in Figure 4. The excitation wavelength was 633 nm, which was pre-resonant with the FUC transition dipole, and thus gave Raman scattering signals above the long-wavelength tail of the fluorescence. The 10 wt% aq. PVA solution with 10 μM FUC was spin-coated on RPE plates, resulting in a dried PVA film with 550 nm in thickness embedded FUC with an effective molar concentration of ~100 μM. The gold(I)/halide treatment intensified the RPE activity of the CSS-protected AgNIs by about a few to ten times in both Raman scattering and fluorescence. The final EF of the RPE plate with the gold(I)/halide bath treatment could be reached up to $2\times10^7$ in the Raman spectrum of R6G and 170 in the fluorescence spectrum of FUC with both the excitation of 532 nm.

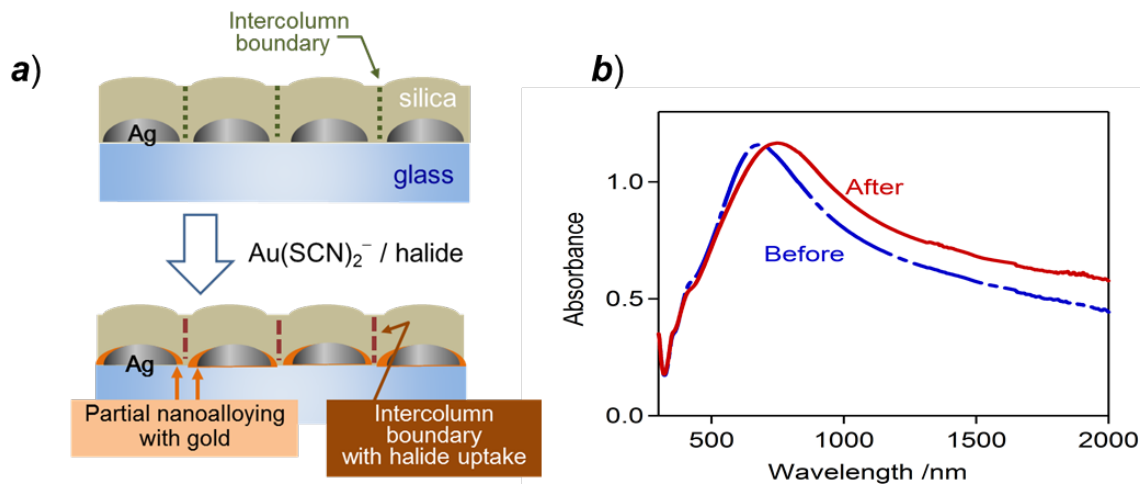

**Figure 3.** Effect of the gold(I)/halide bath treatment on an RPE plate. (a) Schematic illustration of the probable effects of the gold(I)/halide bath treatment at elevated temperatures above 60°C. The vertical broken lines are for CSS intercolumn boundaries. See the discussion for more details. (b) Extinction spectra of the CSS-protected AgNIs taken before and after the gold(I)/halide bath treatment.

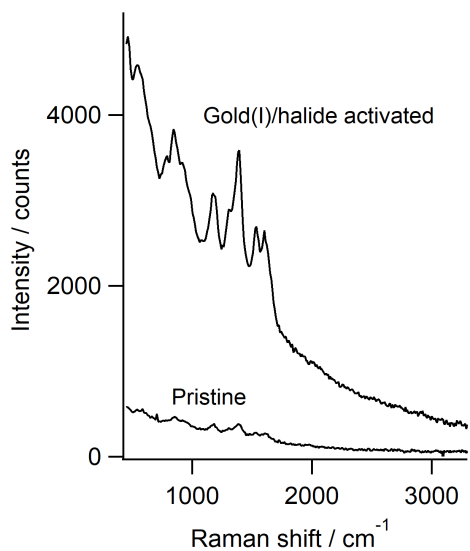

**Figure 4.** The effect of gold(I)/halide treatment on the RPE-enhanced emission spectrum. The RPE-enhanced emission spectra with overlapping signals of Raman scattering and fluorescence of FUC embedded in PVA films on the RPE plates were observed with and without the gold(I)/halide treatment at 633-nm excitation. The effective molar concentration of FUC was 100 µM in PVA film with 550 nm in thickness.

### 3.3 Enhancement characteristics of RPE

To characterize the enhancement effect of RPE, we first evaluated the effect of two opposite optical configurations, i.e., front- or rear-side excitation and on the same side collection of emission, as shown in Figure 5a. The RPE-enhanced spectra exhibited analogous signal intensities for the two opposite optical configurations for both Raman scattering and fluorescence, as shown in Figures 5b and 5c. The fact that the front/rear excitation and signal collection resulted in the same spectral intensity despite the 90% light attenuation by the dense AgNI array of the RPE plate suggested that LSPR of the AgNIs was intimately involved both in the excitation and emission processes.

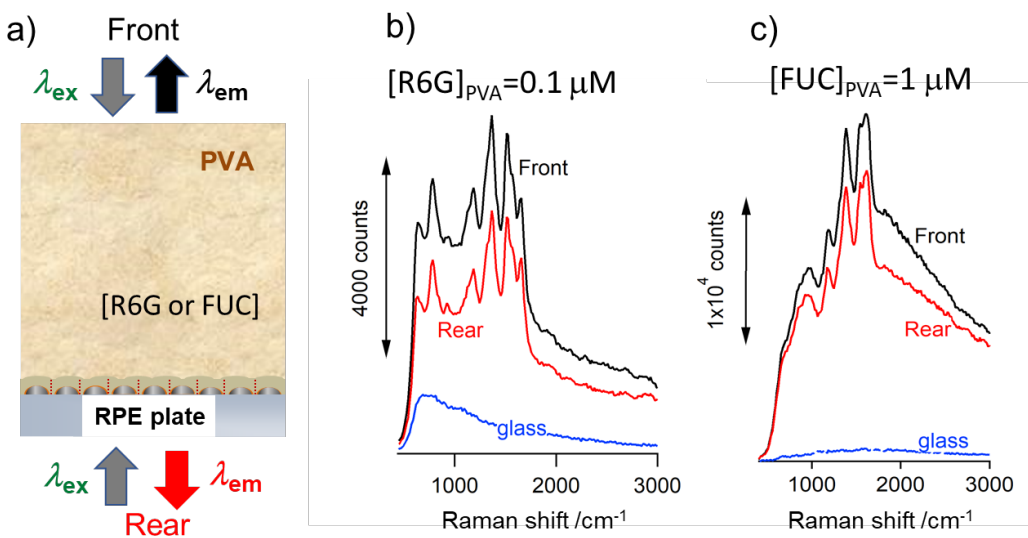

**Figure 5.** RPE-enhanced Raman and fluorescence spectroscopies with two opposite optical configurations. (a) A schematics of the two opposite optical configurations, i.e., front- or rear-side excitation and on the same side collection of emission. RPE-enhanced emission spectra with overlapping Raman scattering and fluorescence signals of (b) R6G embedded with 0.1 μM and (c) FUC with 1 μM embedded in the PVA films at 532-nm excitation. A reference spectrum taken for the same concentration of each PVA-embedded analyte molecule on a transparent slide glass is shown in blue.

We next investigated EFs in RPE-enhanced Raman and fluorescence spectroscopies. Suppose RPE occurs solely by the electric field enhancement originating from AgNIs or CSS nanostructures of the RPE plate; the EFs for Raman scattering and fluorescence should be consistent because of linear dependence on electric field strength. However, a far more substantial enhancement in Raman scattering than fluorescence was observed, as shown in Figures 5b and 5c. Furthermore, the EFs in fluorescence were also different depending on molecular spectroscopic properties. In R6G, the fluorescence EF relative to the reference signal measured on glass was at most ~4, as shown in Figure 5b. R6G acts without RPE as a high fluorescence quantum yield (> 0.95) fluorophore[34] both in solution and in the PVA-embedded state. Considering its original fluorescence quantum yield close to unity, suppose that this extra enhancement arose mainly from an increase in the excitation efficiency, in other words, effective absorption cross-section. If so, the fluorescence of FUC would have to show EF of ~4 likewise. However, it was ~20-fold which was much higher than that of R6G in the given excitation condition, as shown in Figure 5c. FUC has a small fluorescence quantum yield compared with R6G, even in the PVA matrix. These dependencies of EFs indicated that RPE was not just an electric field enhancement.

We also investigated the molecular concentration dependency for RPE. Figure 6a shows a FUC concentration series of RPE-enhanced emission spectra in a logarithmic intensity scale with overlapping Raman and fluorescence spectra taken at the excitation wavelength of 532 nm. In the lower concentrations (<10 μM), the Raman and fluorescence intensities increased almost proportional to the molecular concentration in both the front- and rear-side measurements. In contrast, in higher concentrations (> 100 μM), nonlinear dependence of the emission spectra on molecular concentration was observed. Namely, the Raman signals became invisible in both configurations in the logarithmic intensity scale, and the fluorescence signal was saturated in the rear-side measurement. This means that the equivalence between the two opposite optical configurations, characteristic of RPE, broke up in the high concentration regime (cf. Figure S3 for the R6G series of emission spectra given in linear intensity scale).

Figure 6b shows how the intensity of a representative Raman band of FUC (1383 cm$^{-1}$) changed with the molecular concentration for the rear-side measurement. In the lower concentrations (< 10 μM), the Raman intensity increased almost proportional to the molecular concentration, whereas it was maintained or decreased even with the increase of molecular concentration in the higher concentrations (>10 μM). These results indicated that effective and simple signal enhancement was obtained under low molecular concentration conditions in RPE for both Raman and fluorescence spectroscopies. Furthermore, under high molecular concentration conditions, a complex optical process might be involved in the RPE process.

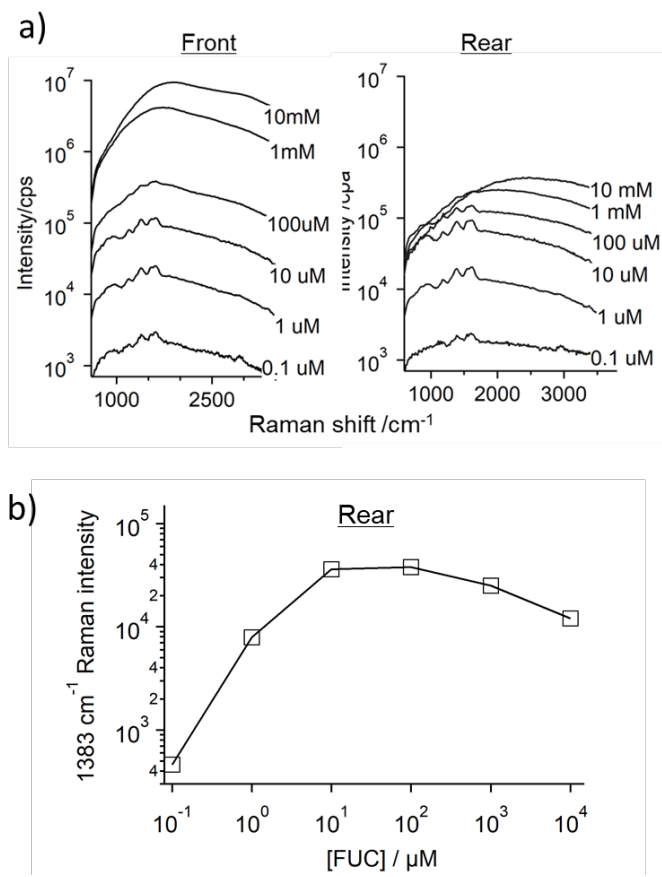

**Figure 6.** Molecular concentration dependency for RPE in a rigorous resonance condition. (a) A wide range of FUC concentration series of RPE-enhanced emission spectra with overlapping Raman and fluorescence spectra by the front- and rear-side measurements at 532-nm excitation. FUC was embedded in a 4-µm-thick PVA film. (b) FUC concentration dependency of Raman intensity for the 1383 cm$^{-1}$ obtained with the rear-side measurement.

The Raman enhancement by RPE was also strongly affected by excitation wavelength. Since FUC had an absorption peak at around 550 nm, the results with the excitation wavelength of 532 nm shown in Figure 6 were in an almost rigorous electronic resonance condition of the excitation wavelength with the FUC molecular transition dipole. In a pre-resonance condition with the excitation wavelength of 633 nm, in which the excitation photon energy was near but lower than the absorption band of the FUC molecular transition dipole, the Raman EF by RPE was still large but decreased by an order of magnitude from the rigorous resonant case at the wavelength of 532 nm, as shown in Figure 7a. Conversely, the proportional Raman enhancement against concentration was observed up to 100 μM for the excitation wavelength of 633 nm, which was a much higher concentration than the corresponding upper limit to sustain such linearity for the excitation wavelength of 532 nm (~10 μM). In a further off-resonance condition of the excitation wavelength of 785 nm (Figure 7b), we observed noticeable enhanced Raman signals only in the considerably high FUC concentration above 100 μM, where the enhanced Raman intensity continued to increase with concentration. These results suggested that RPE was modulated by the electronic resonance between the excitation light and analyte molecules, which significantly affected the enhancement properties of RPE.

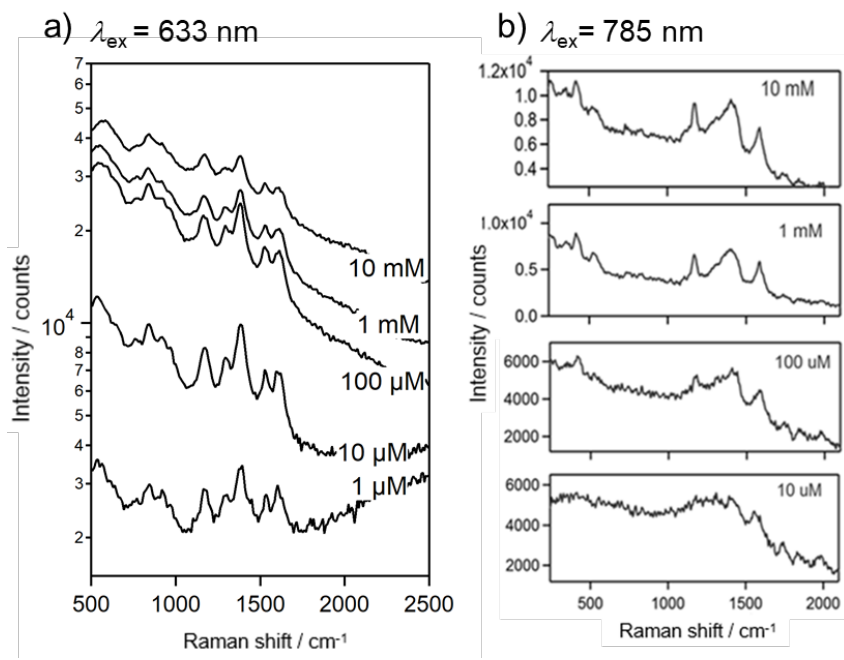

**Figure 7.** Molecular concentration dependency for RPE in pre- and off-resonance conditions. FUC concentration series of RPE-enhanced emission spectra with overlapping Raman and fluorescence spectra taken (a) at the wavelength of 633 nm in a pre-resonance condition (shown in logarithmic intensity scale) and (b) at the wavelength of 785 nm in an off-resonance condition.

**3.4 RPE-Enhanced Fluorescence Biosensing for Live Cells**

For the application of the RPE in fluorescence biosensing, we evaluated the RPE-enhanced fluorescence detection capacity of $Ca^{2+}$ oscillation of live cells. Precoating of the RPE plate with protein Matrigel led to a healthier growth of HeLa cells as like a general Matrigel-coated slide glass, as shown in Figure 8a. We employed fluo3-AM dyes as a cytosolic $Ca^{2+}$ sensor (Figure 8b), which has been extensively used as one of the most potent $Ca^{2+}$ indicators of the current biosensing technology.[35] We used a wide-field fluorescence microscope at the excitation wavelength of 480 nm.

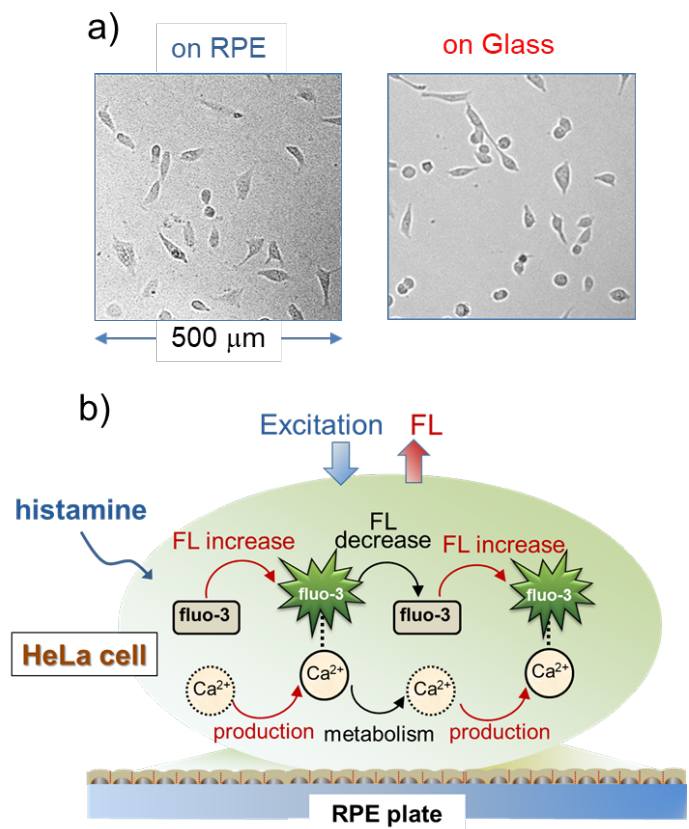

**Figure 8.** Biosensing of live cells on an RPE plate. (a) Each of the optical micrographs of HeLa cells cultured on a Matrigel-coated RPE plate and glass. (b) Schematics of $Ca^{2+}$ oscillation induced by histamine and monitored with fluo3-AM as a biosensor of $Ca^{2+}$.

The time response curves of $Ca^{2+}$ levels of several cells induced by histamine stimulus at the fluo3-AM loading level of 0.1 μM in the extracellular medium are shown in Figure 9a. The fluorescence signal of the HeLa cells on the RPE plate was significantly enhanced by comparing with that on the glass plate. This significant enhancement by RPE led to the clear visualization of $Ca^{2+}$ oscillation of HeLa cells induced by histamine.

The significant enhancement effect of RPE was further illuminated using two quantitative terms in the time response curve; the initial fluorescence spike intensity ($I_s$) by the histamine stimulus and the number of $Ca^{2+}$ signal oscillations ($N_{osc}$). $I_s$ serves as a measure of the initial transient rise in the concentration of the complex between fluo3-AM and $Ca^{2+}$ in each cell. $N_{osc}$ is an important statistical measure of the intrinsic signaling dynamics of $Ca^{2+}$. The distributions of $I_s$ and $N_{osc}$ are shown in Figure 9b. We employed HeLa cells exhibiting $I_s$ over the noise level for analysis. As the fluorescence intensity of $I_s$ became stronger by RPE, $N_{osc}$ also tended to increase. This might be because the $Ca^{2+}$ oscillation following the initial spike would be buried in noise on the glass plate, while the enhanced $Ca^{2+}$ oscillation signal on the RPE plate was adequately observed. These results provided the proof-of-principle demonstration of the feasibility and efficacy of RPE in fluorescence biosensing for live cells.

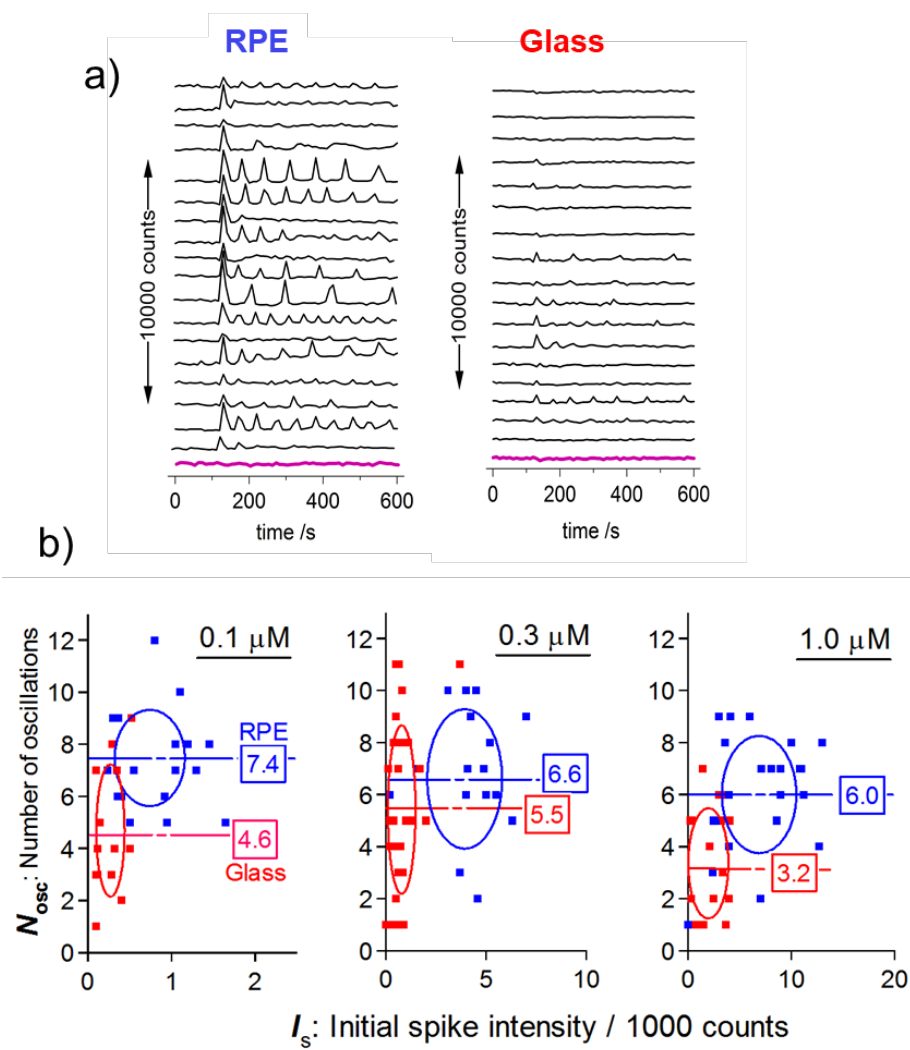

**Figure 9.** Fluorescence observation of Ca$^{2+}$ oscillation of HeLa cells induced by histamine. (a) Temporal fluorescence signal behavior of Ca$^{2+}$ levels of HeLa cells on RPE or glass plates. The Ca$^{2+}$ oscillation was induced by 1 µM of histamine. The Ca$^{2+}$ indicator of fluo3-AM was loaded at the concentration of 0.1 µM in the extracellular medium. The pink lines show the instrumental noise level. (b) Scatter plots of the initial fluorescence spike intensity ($I_s$) and the number of oscillations ($N_{osc}$) for the three different levels of fluo3-AM loading. The time window for counting the $N_{osc}$ was set with 480 s from the initial spike of Ca$^{2+}$ oscillation. Numbers in boxes indicate the average of $N_{osc}$. The center, major diameter, and minor diameter of the ellipses correspond to the mean and the two times of standard deviations, each of $I_s$ and $N_{osc}$.

### 3.5 Enhanced Tissue Raman Imaging

We show next how RPE works in histological Raman imaging. A tissue section of an esophagus with esophageal adventitia of a Wistar rat was attached onto an RPE plate, as shown in Figure 10a. We used a line illumination confocal Raman microscope operating at the excitation wavelength of 532 nm.

Raman spectra of the respective tissue domains distinguished in the optical image are shown in Figure 10b. The distinct Raman spectra of vagus nerves, adipose tissues, blood vessels, and smooth muscles were clearly observed by using the RPE plate (red spectra). Whereas on slide glass (black spectra), we could obtain discernable Raman signals for only adipose tissues and blood vessels by the high concentration of fatty acids in the adipose tissues and by the resonant Raman effect of hemoglobin in the blood. The EFs of the RPE-enhanced Raman signals derived from other than fatty acids and hemoglobins were about 10$^4$ or higher, judging from the noise level of the Raman spectrum on the glass. The EFs of the RPE-enhanced Raman signals derived from fatty acids and hemoglobins were about 10.

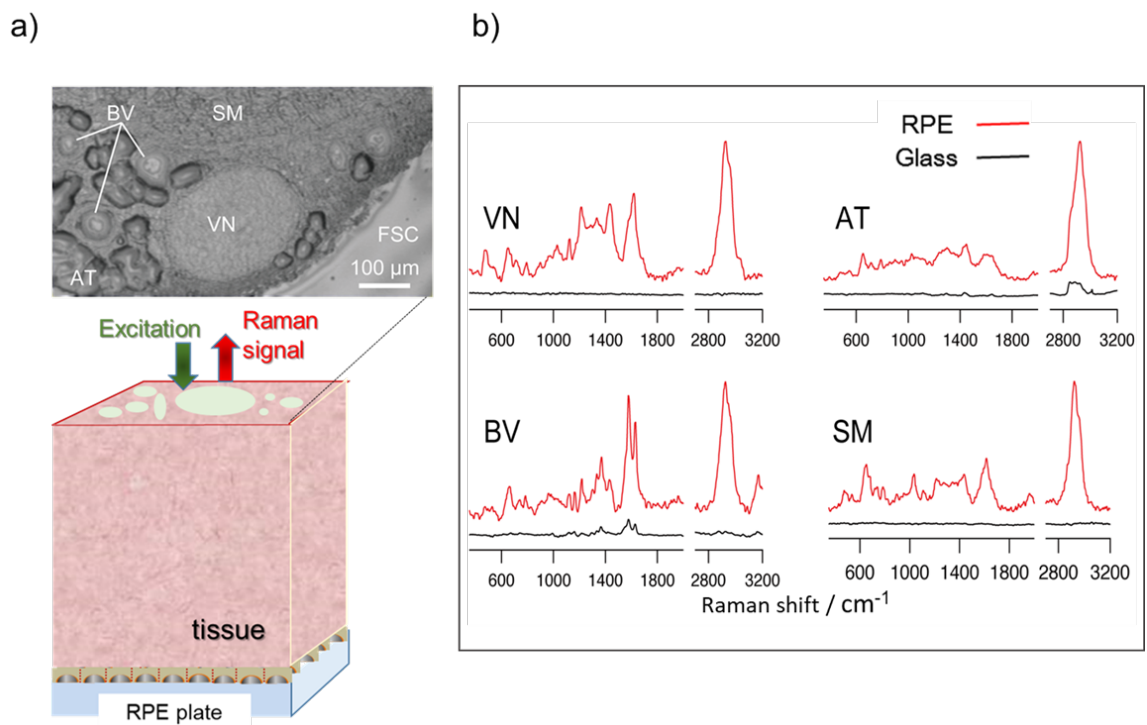

**Figure 10.** RPE-enhanced Raman spectroscopy of esophagus with esophageal adventitia of a Wistar rat. (a) A white-light image and schematics of a tissue section of an esophagus with esophageal adventitia of a Wistar rat attached on an RPE plate. Each distinguishable tissue domain is marked by its respective name in the white-light image. VN, vagus nerve; AT, adipose tissue; BV, blood vessel; SM, smooth muscle; and FSC, frozen section compound. (b) Raman spectra taken on the RPE plate (red) and on glass (black) for each tissue domain.

We also evaluated the tissue imaging capability of RPE, as shown in Figure 11. RPE-enhanced Raman images represented unique patterns that highlighted tissue domains. Table 1 summarizes this relationship along with the most probable vibrational assignments. In particular, the two images at 1431 and 1450 cm$^{-1}$ showed a sudden turn of contrast from that in favor of vagus nerves to that of adipose tissue, despite the minor difference in Raman shift. This suggested that the RPE-enhanced Raman imaging sensitively reflected a subtle difference in the Raman shift associated with δ(CH) between the vagus nerve and adipose tissue domains.

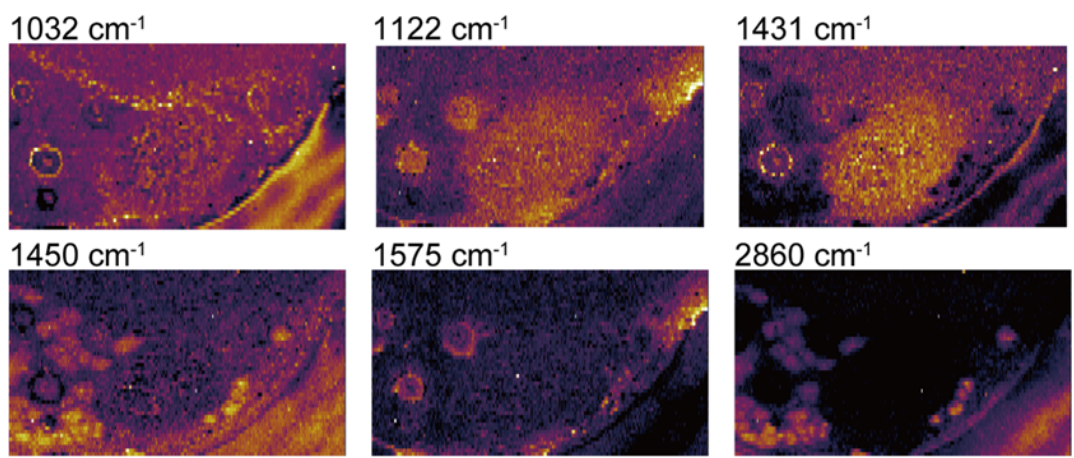

**Figure 11.** RPE-enhanced Raman images of 6 individual Raman bands.

**Table 1.** Assignment of Raman bands that highlighted tissue domains.

| Raman band / cm$^{-1}$ | 1032 | 1122 | 1431 | 1450 | 1575 | 2860 |
|---|---|---|---|---|---|---|
| Predominant domain highlighted | FSC | VN | VN | AT | BV | AT, FSC |
| Assignment [Source substance] | ν(C-O-C) [PEG] | ν(CN) [Protein] | δ(CH) [Protein] | δ(CH) [Lipid] | – [Fe hem] | ν(CH) [Lipid, PEG] |

CT, connective tissue; FSC, frozen section compound; VN, vagus nerve; AT, adipose tissue; BV, blood vessel; PEG, polyethylene glycol.

### 3.7 Mechanism Discussion on RPE

In this study, we have demonstrated RPE by AgNIs with CSS overlayer with more than 100 nm in thickness in fluorescence and Raman spectroscopy. The physical and chemical protection by the CSS layer could lead to reducing the mutual impact between analyte molecules and metal nanostructures. The CSS-protected AgNIs enabled a long-range plasmonic enhancement of fluorescence and Raman scattering, which we referred to as RPE. We found that the gold(I)/halide bath pre-treatment of the RPE plate intensifies the RPE activity. The enhancement effect by RPE has been confirmed with various molecules, including biological specimens.

The conventional plasmonic enhancement of fluorescence and Raman scattering has been realized by metal nanostructures with sufficient proximity of the metal surfaces to analyte molecules. Whereas in the present study, we found a possibility for enhancement without such a range limitation, namely, RPE. We obtained comprehensive experimental support for RPE, such as 1) the markedly upgraded RPE performance after the gold(I)/halide bath treatment, 2) the excitation wavelength dependency manifesting a huge difference in the Raman EF between the resonant and off-resonant excitation conditions, 3) the analyte molecular concentration dependency uncovering a strong preference for a low analyte concentration, and 4) the correlation between the front- and rear-side optical configurations suggesting the intimate role of LSPR both in the excitation and emission processes.

Further confirmation of RPE was performed with the biological samples, demonstrating enhanced fluorescence imaging of $Ca^{2+}$ signaling of living HeLa cells and enhanced Raman imaging of the histology of a rat esophageal tissue. We thereby found also that the RPE plate could provide the spatial resolution that was sufficient for identifying and distinguishing the cells and tissues. This novel enhancement modality of fluorescence and Raman spectroscopy will provide an important insight in terms of the novel mechanism of plasmonic enhancement and for extending the applications of plasmon-enhanced fluorescence and Raman spectroscopy.

As for a possible mechanism of RPE, the resonant coupling between the metal nanoislands and the analyte molecules or the dipolar coupling between the giant plasmon dipole of LSPR and the molecular transition dipole of the analyte may play an important role, as indicated by the correlation between the front- and rear-side optical configurations. This possible mechanism of resonant coupling can also be inferred from the molecular concentration dependency and the wavelength dependency. At higher molecular concentration conditions, the intermolecular distance of analyte molecules becomes shorter, and the Förster-type transition dipole-dipole coupling between the analyte molecules may frequently occur by resonantly exciting the molecular dipole with light.[36] Under these conditions, the resonant coupling between the metal nanoislands and the analyte molecules, if any, would be disturbed. This effect may reflect in the molecular concentration dependency in the resonant case shown in Figure 6. At the concentration of 100 μM or higher of FUC, the average intermolecular distance was expected to be 25 nm or less, resulting in the random pairing of molecules within the Förster distance of ~6 nm would quite frequently occur. As a result, the concentration-dependent Raman enhancement might be no longer observed in the higher molecular concentration condition, possibly due to the disturbance of the resonant coupling between the metal nanoislands and the analyte molecules. In the off-resonant excitation condition, the molecular transition dipole becomes very small, and the intermolecular coupling may become insignificant compared to the plasmon-molecule coupling, resulting in the Raman enhancement according to the molecular concentration was observed even in the relatively higher molecular concentration conditions, as shown in Figure 7. However, the minor molecular transition dipole involved in the plasmon-molecule coupling makes the Raman EF decrease accordingly, as demonstrated in Figures 6 and 7.

As for PVA, we used it as a polymer matrix to embed R6G and FUC dyes. We nevertheless observed no significant Raman signals associated with PVA under any excitation conditions. This is also consistent with the expected mechanism of RPE since PVA is transparent for wavelengths down to the far UV region affording minor transition dipoles by visible light excitation. PVA thus serves as an RPE-inert polymer matrix to disperse arbitrary analyte molecules of interest therein.

In the case of RPE-enhanced fluorescence biosensing of HeLa cells, fluorescent molecules incorporated in the host cells gain a large transition dipole in the given excitation condition and expect to undergo a strong enough coupling with metal nanoislands to bring the fluorescence enhancement. However, an excess amount of the fluorescence probes likely results in the failure of RPE due again to the interference by the intermolecular coupling.

## 3.8 Role of CSS

The aforementioned resonant coupling between metal nanoislands and analyte molecules is not mediated over long distances in common bulk materials. Thus, CSS is likely to act as a structure mediating long-range resonant coupling between the metal nanoislands and the analyte molecules.

The CSS divided by the directional grain boundaries offers a strongly anisotropic solid medium. This anisotropy of the CSS expects to play an essential role in mediating the long-range resonant coupling. One evidence is the further enhancement of RPE by the gold(I)/halide bath treatment. The gold(I)/halide bath treatment might decorate the intercolumn boundaries of CSS with the strongly polarizable halide ions (see Figure 3a). The resultant dielectric modification of the intercolumn boundaries of CSS would facilitate the long-range resonant coupling between metal nanoislands and analyte molecules.

If this hypothesis is correct, a similar promotion effect of the RPE could also be obtained with a gold-free halide bath treatment. Actually, we observed that plain halide bath treatment also brought about a remarkable increase in the RPE activity (data not shown). However, halide ions taken up into the CSS layer, especially at the considerably high processing temperatures, simultaneously caused a rapid corrosive halogenation of the AgNIs to deprive them of the RPE activity. The gold(I)/halide bath treatment allowed us to circumvent this problem, which we attributed to a partial nanoalloying of the AgNIs by the $Au(SCN)^{2-}$ ions protecting against the corrosive halogenation by halide ions, as a similar manner of the process of gold latensification in the silver halide photography.

### 3.10 Bioanalytical Impact of RPE

The present results suggest that RPE serves as a useful potential platform for enhanced biophotonic analyses. The most significant advance of RPE is the sizable enhancement of the Raman scattering and fluorescence signals without the proximity of metal nanostructures and the biological specimens. The CSS layer thicker than 100 nm serves as a robust protection layer for the vulnerable AgNIs to survive highly corrosive solution environments that are employed in biological experiments. It also becomes possible to cultivate live cells atop the CSS protection layer of the RPE plate in the standard manner employing saline as a culture medium. It can also easily attach thin-sliced biological tissues onto the RPE plate. The analyte molecules placed at least ~100 nm apart from the metal surface no longer suffer from short-range electronic interference by the metal. Therefore, RPE enables fast and sensitive molecular sensing in a more biocompatible environment.

In addition, not only these advantages but also the following effects can be obtained. In the case of tissue Raman imaging, RPE may work with a resonate coupling between metal nanostructures and analyte molecules, resulting in sensitive reflection of the electronic properties of the respective tissue materials in the Raman EFs by RPE. Consequently, unique Raman imaging reflecting both the vibrational and electronic states of the analyte molecule can be performed, as shown in Figure 10.

In the case of fluorescence biosensing for live cells, the unique feature of RPE in that it strongly favors a low analyte concentration affords an extra advantage of RPE because one always has to care about the interference effects by the external fluorescence probes. In fact, an adverse effect of fluo3-AM load might appear at the highest fluo3-AM loading level of 1.0 μM. The fluorescence intensities of fluo3-AM of a substantial portion of the HeLa cells yield high enough (>1000 counts) to analyze $Ca^{2+}$ oscillation in both the RPE plate and slide glass. However, the number of oscillations tended to be lower than that at the lower concentration conditions in both the RPE plate and slide glass. This may be understood in terms of the positive feedback mechanism of cells in the regulation of intracellular $Ca^{2+}$. In this mechanism, the initial $Ca^{2+}$ release from the internal $Ca^{2+}$ stores, corresponding to the first spike, promotes the $Ca^{2+}$ influx corresponding to the subsequent oscillations. The sequestration of free $Ca^{2+}$ by complexing with excess fluo3-AM in the overdosed cells results in insufficient initial $Ca^{2+}$ rise, thus impairing the intracellular dynamics of the positive feedback loops. Whereas on the RPE plate at the low concentration conditions, the weakened fluorescence signals were sufficiently amplified with the less adverse effect of fluo3-AM load on $Ca^{2+}$ signaling, enabling the $Ca^{2+}$ oscillations to be counted more accurately.

## 4. CONCLUSION

We demonstrated RPE by a dense random array of AgNIs with a CSS overlayer of more than 100 nm thickness. The present RPE plate affords practical advantages for potential biophotonic analyses. RPE plate is realized just by sputtering and chemical immersion processes, enabling large-area RPE plates to be easily fabricated. Biocompatible structure with CSS provides a more biocompatible molecular detection environment to reduce the mutual impact between analyte molecules and metal nanostructures. It promises high analytical sensitivity and a short acquisition time for biophotonic measurements without any instrumental modification or certain sample manipulation. We thus anticipate that RPE will advance to versatile analytical tools in chemistry, biology, and medicine.

## ASSOCIATED CONTENT

**Supporting Information**

The following files are available free of charge.

Comparison of 633-nm excited Raman spectra on a slide glass and an RPE plate (Figure S1), adhesive tape test to verify RPE-enhanced Raman signals (Figure S2), a strong effect of the concentration of R6G molecules embedded in a PVA film coated on the RPE plate on the series of emission spectra measured in the front- and rear-side configurations of excitation and signal collection (Figure S3) (PDF)

## AUTHOR INFORMATION

**Corresponding Author**

*Takeo Minamikawa – Division of Interdisciplinary Researches for Medicine and Photonics, Institute of Post-LED Photonics, Tokushima University, Tokushima 770-8506, Japan; orcid.org/0000-0003-3528-7163; Email: minamikawa.takeo@tokushima-u.ac.jp

**Author Contributions**

§ T.M. and M.K. contributed equally to this work. The manuscript was written through the contributions of all authors. All authors have given approval to the final version of the manuscript.

**Funding Sources**

This work was partially supported by PRESTO program (JPMJPR17PC) of Japan Science and Technology Agency (JST), Japan, JSPS KAKENHI (15K12519, 19K22969, 21H01847) from the Japan Society for the Promotion of Science (JSPS), Japan, a research grant from the Research Clusters program of Tokushima University (1802003), Japan.

**Notes**

T.M., Y.H., T.T., Ya.M., R.S., and H.H. declare no competing financial interests. Y.Y. and Yu.M. were employees of Ushio Inc. during this project. Mi.K., Yu.M., Ma.K. and Y.Y have filed patents related to this work.

**ACKNOWLEDGEMENT**

We acknowledge technical support from Mr. K. Hayashi (Kyoto University). We also thank Ms. A. Murakami of Tokushima University for her help in the English proofreading of the manuscript.